\documentclass[10pt,aps,pra,twocolumn,showpacs,superscriptaddress,nobalancelastpage,notitlepage,longbibliography]{revtex4-1}

\usepackage{graphicx,psfrag,color,mathbbol,amsthm,amsmath,amsfonts,amssymb,bm,braket,footmisc,multirow,latexsym,times}

\usepackage[colorlinks]{hyperref}
\hypersetup{
 colorlinks=true,
 citecolor=magenta,
 linkcolor=blue,
 urlcolor=cyan}

\graphicspath{{Figures/}}

\newcommand{\tr}{\text{tr}}
\newcommand{\norm}[1]{
\lVert#1
\rVert}

\begin{document}

\title{Topology by Dissipation: Majorana Bosons in Metastable Quadratic Markovian Dynamics}

\author{Vincent P. Flynn}
%\thanks{vincent.p.flynn.gr@dartmouth.edu}
\affiliation{\mbox{Department of Physics and Astronomy, Dartmouth College, 6127 Wilder Laboratory, Hanover, New Hampshire 03755, USA}} 

\author{Emilio Cobanera} 
%\thanks{cobanee@sunyit.edu}
\affiliation{\mbox{Department of Mathematics and Physics, SUNY Polytechnic Institute, 100 Seymour Rd, Utica, NY 13502, USA }}
\affiliation{\mbox{Department of Physics and Astronomy, Dartmouth College, 6127 Wilder Laboratory, Hanover, New Hampshire 03755, USA}} 

\author{Lorenza Viola}
%\thanks{lorenza.viola@dartmouth.edu}
\affiliation{\mbox{Department of Physics and Astronomy, Dartmouth College, 6127 Wilder Laboratory, Hanover, New Hampshire 03755, USA}} 

\begin{abstract} 
Majorana bosons, that is, tight bosonic analogues of the Majorana fermionic quasi-particles of condensed-matter physics, are forbidden for gapped free bosonic matter within a standard Hamiltonian scenario. We show how the interplay between dynamical metastability and non-trivial bulk topology makes their emergence possible in non-interacting bosonic chains undergoing Markovian dissipation. This leads to a distinctive form of \emph{topological metastability}, whereby
%, consistent with the breakdown of Noether's theorem,
a conserved Majorana boson localized on one edge is paired, in general, with a \emph{distinct} symmetry generator localized on the opposite edge. We argue that Majorana bosons are robust against disorder and identifiable by signatures in the zero-frequency steady-state power spectrum. Our results suggest that symmetry-protected topological phases for free bosons may arise in transient metastable regimes, which persist over practical time scales. 
\end{abstract}

\date{\today}

\maketitle

The discovery of phases of matter exhibiting topological order revolutionized our understanding of many-body quantum systems, by challenging the Ginzburg-Landau theory of local order parameters. Symmetry-protected topological (SPT) phases of free (mean-field) bulk-gapped fermionic matter in equilibrium have been fully classified \cite{ryuRMP}. A pillar of the ensuing bulk-boundary correspondence is the emergence of zero-energy boundary-localized Majorana fermions (MFs) that are robust against symmetry-preserving local perturbations and, thanks to the non-Abelian braiding statistics they can engender, could enable topological quantum computation \cite{KitaevTQC,IntroTQC}. In contrast, a series of no-go theorems rules out the existence of SPT phases and bosonic analogues of MFs -- \emph{Majorana bosons} (MBs) -- in systems described by a gapped, stable quadratic Hamiltonian \cite{qiaoru}. Non-trivial topology may emerge if one allows for gapless phases \cite{topophrev}, strong interactions \cite{Chen2012,Bardyn2012,Ashvin3DTopo}, or instabilities \cite{ClerkPeano,clerkPRX,Decon}. Does this then altogether rule out Majorana physics for non-interacting bosons and the possibility to build insight starting from a single-particle picture?

The search for topological order in non-interacting matter has extended into the realm of open quantum dynamics -- both in a semiclassical limit described by non-Hermitian effective Hamiltonians \cite{NHtopoclass,WangNHSE,SlagerNHBmodes,TopoNHSE,TopoSensor}, and a fully quantum setting described by a quadratic Markovian semigroup \cite{alicki-lendi}. For fermionic matter, the use of engineered dissipation has been shown to provide a compelling paradigm for accessing topologically non-trivial steady states, and significant advances have been made towards uncovering a dissipative bulk-boundary correspondence which links bulk invariants to dissipative Majorana edge modes \cite{diehl,bardynNJP}. By furnishing almost conserved operators, these long-lived modes can dramatically alter the transient dynamics. They elicit correlation and relaxation times that may diverge with system size \cite{GarrahanPRB,Vernier}, reminiscent of \emph{metastability} \cite{GarrahanPRL}. For bosons, the situation remains far more opaque. On the one hand, connections have been made between certain bulk topology and directional amplification of an input driving field \cite{PorrasPRL,NunnenkampNat,NunnenkampDisorder,PorrasIO}; on the other hand, there has yet to be any identification of the analogous MBs that are so fundamentally tied to SPT phases. 

In this Letter we establish the existence of MBs for systems described by quadratic bosonic Lindbladians (QBLs). MBs are linked to non-trivial bulk topology and to a form of \emph{topological metastability} that we argue is unique to bosons. By leveraging tools from pseudospectral analysis \cite{Trefethen}, we show that, despite the system being dynamically stable for all finite size, unstable behavior may arise in the thermodynamic limit, enforcing anomalously long prethermalization and transient amplification. Metastability is necessary but \emph{not} sufficient for MBs, however: topological metastability, featuring MBs, additionally requires that a winding number around zero be non-vanishing. Consistent with the fact that symmetries and conserved quantities are generally independent  
%breaking of Noether's theorem 
in open dynamics  \cite{Baum,ProsenSpin,Victor,GoughNoether}, we find that a MB pair generically comprises a conserved zero mode (ZM) localized on one edge, and a symmetry generator, localized on the other -- together forming a split bosonic degrees of freedom. The symmetry is responsible for a continuum of \emph{quasi-steady states} that can lead to unexpected features, like persistent non-Gaussianity. We identify simple models and show how MBs lead to distinctive signatures in experimentally accessible steady-state power spectra. Thanks to the robustness of pseudospectra against perturbations, our results carry through in the presence of weak disorder. 

{\em Quadratic bosonic Lindbladians.---} Consider the Lindblad master equation $\dot{\rho}(t) = \mathcal{L}(\rho(t))$ for a density operator $\rho(t)$ on an $N$-mode bosonic Fock space. The observables \(A\) carry the time dependence in the adjoint ($^\star$) Heisenberg picture, so that  
\begin{eqnarray}
\label{LH}
\dot{A}(t) = \mathcal{L}^\star(A(t)) \equiv i[H,A(t)] + \mathcal{D}^\star(A(t)),\quad t\geq t_0, \;
\end{eqnarray}
where the Hamiltonian $H=H^\dagger$ need not be bounded below and the dissipator $\mathcal{D}^\star$ is bilinear in the Lindblad operators $\{L_k\}$ \cite{alicki-lendi}. Let $\Phi \equiv [a_1,a_1^\dag,\ldots,a_N,a_N^\dag]^T$ be a Nambu array of bosonic annihilation, creation operators. Then, $H \equiv \frac{1}{2}\Phi^\dag \mathbf{H} \Phi$ is a quadratic bosonic Hamiltonian associated to the matrix \(\mathbf{H}=\bm{\tau}_1 \mathbf{H}^* \bm{\tau}_1\). Hereafter, $\bm{\tau}_j\equiv \bm{1}_N\otimes \bm{\sigma}_j$, in terms of the Pauli matrices. As the Lindblad operators $L_k\equiv \sum_{j=1}^{2N} \ell^k_j \Phi_j$, $\ell^k_j \in {\mathbb C}$, are linear in the bosonic operators, we have $\mathcal{D}^\star(A) = \sum_{i,j=1}^{2N}\mathbf{M}_{ij} \left(\Phi_i^\dag A\Phi_j - \frac{1}{2}\left\{\Phi^\dag_i\Phi_j,A\right\}\right),$ with $\mathbf{M}_{ij}\equiv \sum_k (\ell_i^k)^* \ell_j^k$ a positive-semidefinite matrix. 
The equations of motion of the linear forms \(\Phi(t)\) and the quadratic forms $Q_{ij}(t) \equiv (\Phi_i\Phi_j^\dag)(t)$ are
\begin{eqnarray}
\dot{\Phi}(t) &=& -i \,\mathbf{G} \Phi(t),\quad 
\label{PhiEOM} \\ 
\dot{Q}(t) &=& -i\left( \mathbf{G}Q(t) - Q(t) \mathbf{G}^\dag\right) +\bm{\tau}_3 \mathbf{M} \bm{\tau}_3 ,
\label{QEOM}
\end{eqnarray}
with the \emph{dynamical matrix} ${\bf G} \equiv {\bm{\tau}}_3 {\bf H} -\frac{i}{2}{\bm{\tau}}_3\left( {\bf M} - {\bm{\tau}}_1 {\bf M}^T {\bm{\tau}}_1\right)$. The \emph{rapidity spectrum}, $\sigma(-i\mathbf{G})$, is the set of eigenvalues of $-i\mathbf{G}$.  Since $\mathbf{G} = -\bm{\tau}_1 \mathbf{G}^* \bm{\tau}_1$, we have $\sigma(-i\mathbf{G})= \sigma(-i\mathbf{G})^*$. 

A quantum system is \textit{dynamically stable} if the expectation value of an arbitrary observable in any state is bounded for all $t\geq t_0$. For a QBL, a sufficient condition for stability is that \(-i\mathbf{G}\) is a Hurwitz matrix, that is, $\sigma(-i\mathbf{G})$ is bound to the open left-half plane. 
In this case, there exists a unique, globally attractive, Gaussian steady state $\rho_\text{ss}$, ${\cal L}( \rho_\text{ss})=0$,  which is completely determined by the expectation values $\braket{\Phi}_\text{ss} = 0$ and $\braket{Q}_\text{ss} = \mathbf{Q}_\text{ss}$ \cite{Gardiner,Teretenkov}. When, in addition, $\mathbf{G}$ is diagonalizable, $\mathcal{L}$ is as well and its spectrum follows from $\sigma(-i\mathbf{G})$ \cite{prosen3fermion,prosenfermspecthm,prosen3boson}. The convergence to the steady state is exponential, with an asymptotic rate determined by the \emph{spectral gap}, $\Delta_\mathcal{L}\equiv |\max\,\text{Re}(\sigma(\mathcal{L})\setminus\{0\})|= |\max\,\text{Re}(\sigma(-i\mathbf{G}))|$. 
More precisely, the worst-case distance $d_\text{max}(t)$ from $\rho_\text{ss}$ satisfies $d_\text{max}(t) \equiv \sup_{\rho(0)}\norm{\rho(t)-\rho_\text{ss}}\leq K e^{-\Delta_\mathcal{L}t},$ with $K$ independent of time. The minimum time it takes for $d_\text{max}(t)$ to fall below a pre-determined accuracy $\delta>0$ is the \textit{mixing time} $t_{\text{mix}}(\delta)$ of the semigroup generated by \(\mathcal{L}\) \cite{alicki-lendi}. 
 
We focus on one-dimensional, bulk translation-invariant QBLs. There are three basic configurations: periodic (PBCs), open BCs (OBCs, two terminations), and semi-open BCs (infinite system with one termination). With hindsight, an infinite system without terminations is well described as the limiting case of PBCs. The dynamical matrices of these QBLs are known as block-Toeplitz or circulant matrices for OBCs and PBCs, respectively \cite{BottcherBook,JPA}. Translation invariance implies the rapidity spectrum for PBCs form closed curves in ${\mathbb C}$ as $N\rightarrow \infty$. We will rely on two model QBLs for illustration. The Hamiltonian is the bosonic Kitaev chain (BKC) \cite{clerkPRX,Decon}, 
\begin{equation}
\label{BKCHam}
H_{\text{BKC}}\! = \frac{i}{2}\sum_{j=1}^{N-1}\left( J a_{j+1}^\dag a_j +  \Delta a_{j+1}^\dag a_{j}^\dag \right) + \frac{i\mu}{2} \sum_{j=1}^{N}(a_j^\dag)^2 +\text{H.c.}, 
\end{equation}
where $J\geq \Delta\geq 0$ are nearest-neighbor hopping and non-degenerate parametric amplification amplitudes, and $\mu \in\mathbb{R}$ is a uniform degenerate parametric amplification strength \cite{Walls}. It is the dissipation mechanisms that separates the two models: 

\noindent
\(\bullet\) {\bf Model 1}: Uniform onsite dissipation, $L_j = \sqrt{2\kappa} \,a_j$ for $j=1,\ldots, N$, with $\kappa > 0$ the uniform damping strength. 

\noindent
\(\bullet\) {\bf Model 2}: Add to the previous model next-nearest-neighbor damping of strength $\Gamma >0$. 

\begin{figure}[t]
\includegraphics[width=0.81\columnwidth]{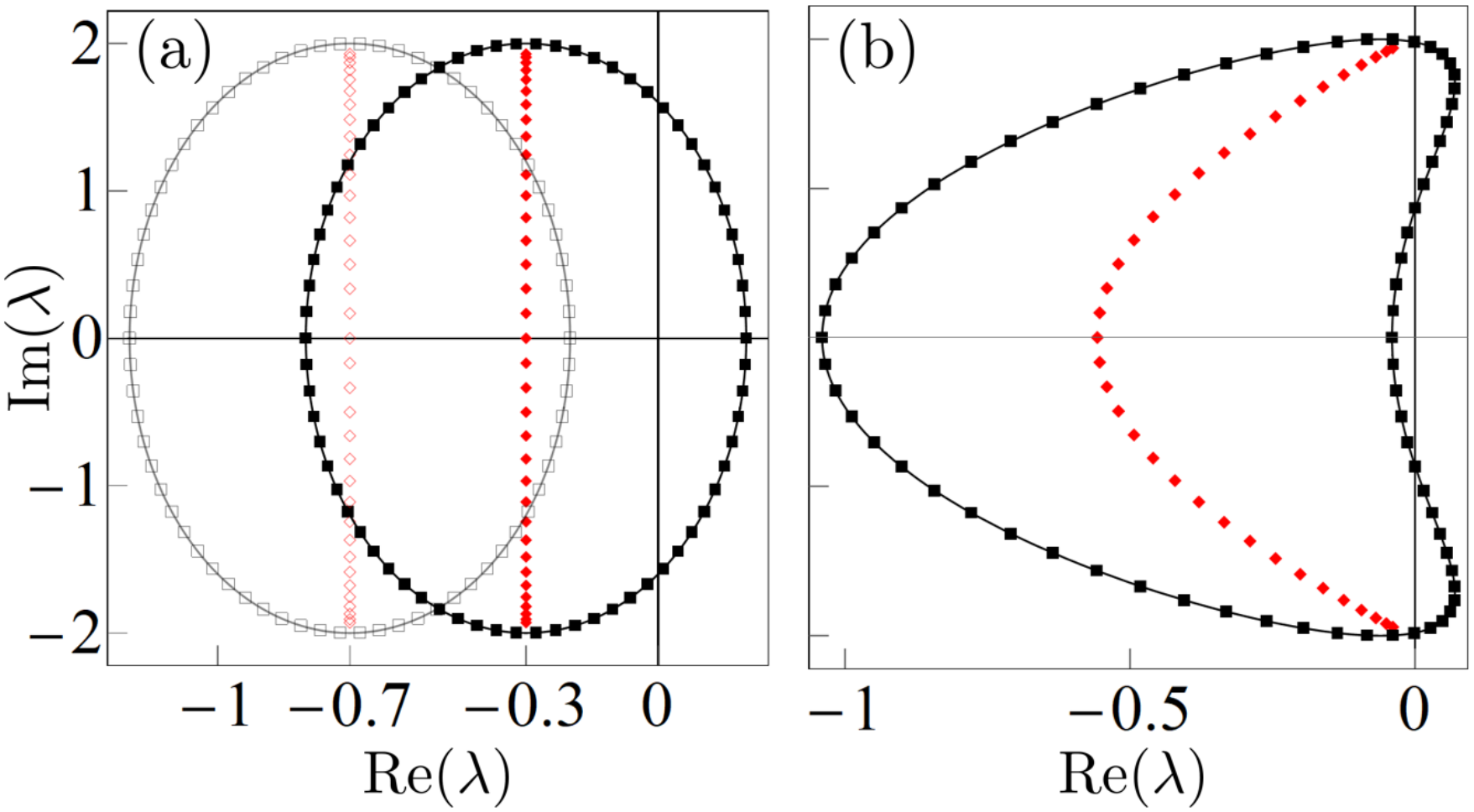}
\vspace*{-3mm}
\caption{(Color online) (a) Rapidity spectrum of Model 1. The filled (open) markers represent the metastable (unconditionally stable) regime, with $\kappa/\Delta=0.6$ ($1.4$). The solid ellipses give the bulk spectrum, whereas the points are the rapidities for PBCs. The points on the vertical lines are the rapidites for OBCs. In both cases $J=2$, $\Delta=0.5$, $\mu=0$, $N=25$.   (b) Rapidity spectrum of Model 2 in the metastable regime. Same parameters as in (a) with $\Gamma=0.12$.  The winding number  of the spectrum} around $\lambda=0$ is \emph{zero} in this case. 
\label{f1_bkcVkidney}
\end{figure} 

{\em Metastable QBLs and $\epsilon$-pseudospectrum.---} Model 1 is dynamically unstable for PBCs whenever $\Delta + |\mu|\geq \kappa$, whereas it is stable for OBCs when $\kappa-|\mu|>0$ (see in Fig.\,\ref{f1_bkcVkidney} for $\mu=0$ and \cite{SM} for details). The dependence on BCs is dramatic for both models and cannot be weakened by increasing $N$. As we will show, any QBL that is (i) dynamically stable for OBCs, and (ii) unstable for PBCs is \emph{metastable}, in the sense that the relaxation proceeds in two steps: a long-lived transient where the system behaves as if it were in an unstable phase, followed by stable asymptotic dynamics. To contrast, we shall call any system that is dynamically stable but not metastable \emph{unconditionally stable}. This metastable behavior is not captured, in general, by the spectral properties of the QBL. The problem is that ${\bf G}$ is non-normal and so the spectral decomposition is fragile; one must drop it in favor of the more robust notion of the $\epsilon$-pseudospectrum \cite{Trefethen}, as also recently used for non-Hermitian Hamiltonians \cite{SatoPP,SatoOkumanew}. 

For a matrix $\mathbf{X}$ and $\epsilon>0$, the {\em $\epsilon$-pseudospectrum} of $\mathbf{X}$ is $\sigma_\epsilon(\mathbf{X}) \equiv \set{\lambda\in\mathbb{C}: \exists \vec{v},\,\norm{\vec{v}}=1,\, \norm{(\mathbf{X}-\lambda \bm{1})\vec{v}}<\epsilon}$. The vectors $\vec{v}$ are the $\epsilon$-pseudoeigenvectors. When $\mathbf{X}=-i\mathbf{G}$ is a Toeplitz matrix, the $\epsilon$-pseudospectrum can be computed by analyzing the winding number of the bulk rapidity bands for PBCs \cite{WN}: if a rapidity band winds around $\lambda \in {\mathbb C}$, then $\lambda$ eventually joins the $\epsilon$-pseudospectrum for any $\epsilon>0$ as the system size increases \cite{SM}.  Thus, the semi-open limit of a metastable system is dynamically unstable. Unlike the spectrum, the $\epsilon$-pseudospectrum is robust against perturbations, in that it scales linearly in the perturbation strength \cite{Trefethen}. Pseudospectra control the transient behavior of a linear dynamical system. Notably, for all $\epsilon>0$, we have that $\sup_{t\geq 0} \norm{e^{\mathbf{X}t}}_2 \geq \sup\,\text{Re}(\sigma_\epsilon(\mathbf{X}))/\epsilon$ \cite{Trefethen}.

The above result may be used to lower-bound the mixing time of an arbitrary QBL. A simple proxy for $t_{\text{mix}}(\delta)$ is the \emph{linear mixing time} $t_\text{lin}(\delta)$, determined by the time it takes for the worst-case distance $d_\text{lin}(t) \equiv \sup_{\braket{\Phi}_0}\frac{\norm{\braket{\Phi}_t - \braket{\Phi}_\text{ss}}}{\norm{\braket{\Phi}_0 - \braket{\Phi}_\text{ss}}}$ to drop below \mbox{$\delta>0$}. From Eq.\,\eqref{PhiEOM}, one finds $d_\text{lin}(t) = \norm{e^{-i\mathbf{G} t}}_2$. If the QBL is metastable, one can show \cite{SM} that for any fixed $\delta<\sup_{t\geq 0}d_\text{lin}(t)$ and $r \geq 0$, there exists $N_{\text{max}}$ such that 
\begin{equation}
t_\text{lin}(\delta, N) >  r  /\Omega , \quad N > N_{\text{max}}, 
\label{tlin}
\end{equation}
where $\Omega >0$ is finite and determined by the dynamical matrix $\mathbf{G}^\text{B}$ of the bi-infinite system. This is consistent with the observation that systems exhibiting the non-Hermitian skin effect can experience abnormally long relaxation without a closing of the spectral gap in the thermodynamic limit and suggests that pseudospectra may explain this generally \cite{WangChiral,UedaLSE,MoriRelaxation}.

\begin{figure}[t]
\includegraphics[width=0.85\columnwidth]{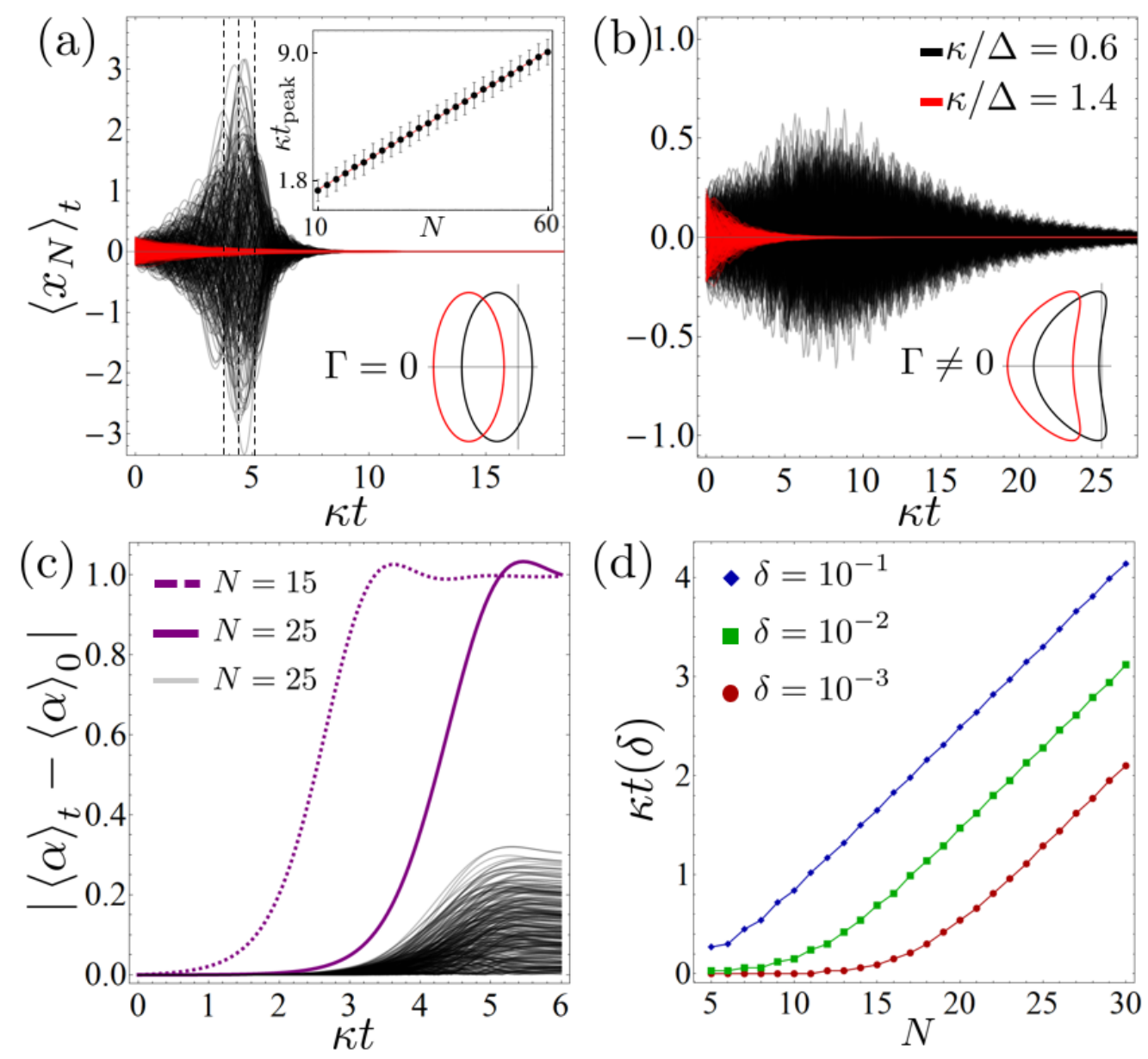}
\vspace*{-2mm}
\caption{(Color online) 
(a) Expectation of $x_N$ vs. time, uniformly sampled over $500$ initial conditions with $\norm{\braket{\Phi}_0}=1$ in the metastable ($\kappa/\Delta=0.6$, black) and stable ($\kappa/\Delta=1.4$, red) regimes of Model 1. The central dashed lines indicate the average peak position, while the left and right dashed lines span a width of two standard deviations about the mean with $N=30$. Upper inset: average peak location for $\braket{x_N}_t$ as a function of $N$, with error bars spanning two standard deviations. Lower inset: rapidity band structure.  (b) Same as in (a) but for Model 2. (c) Thin gray curves: Expectation values of 250 randomly sampled linear observables in the quasi-steady state $\rho_{\theta=1}(t)$  generated by the left-localized MB $\gamma_L^\text{s}$, in the BKC, 
with $\kappa/\Delta=0.6$, $N=25$, $\norm{\vec{\alpha}}=1$. Thick purple curves: The upper bound in Eq.\,\eqref{mUB} for $N=15$ (dashed) and $N=25$ (solid). (d) The time $t(\delta)$ it takes for the aforementioned upper bound to exceed accuracy $\delta$ as a function of $N$. In all plots, $J=2$ and $\Delta=0.5$. }
\label{f2_expamplif}
\end{figure}

Another manifestation of metastability is a time of \emph{transient amplification} that increases with system size. In Fig.\,\ref{f2_expamplif}(a)-(b) we plot the expectation values of the boundary observable $x_N$ over uniformly sampled initial conditions, in the metastable and stable regimes for the two models. In the former regime, $x_N$ is exponentially amplified (while $x_1$ decays, not shown). In the unconditionally stable regime, exponential decay dominates, regardless of spatial location. We also plot in (d) the mean peak location for $\braket{x_N}_t$ in the metastable regime against system size. This quantity, which lower-bounds a generic relaxation time-scale, diverges linearly with $N$.  

\emph{Topological metastability and MBs.---} The MFs of condensed matter physics are the Hermitian zero-energy boundary modes of topologically non-trivial superconductors: e.g., the fermionic Kitaev chain (FKC) \cite{Kitaev} can display two MFs localized on opposite ends of the chain. These modes can be combined into a single (Dirac) fermionic ZM with support only on the boundary. Tight bosonic analogues of MFs do not exist for thermodynamically stable quadratic bosonic Hamiltonians \cite{qiaoru}. Hints of them appear if stability is lifted: e.g., the BKC of Eq. \eqref{BKCHam} with $J=\Delta$ hosts approximate ZMs  \cite{Decon}, 
\begin{equation}
\label{Deconbosoranas}
\gamma_{L} = \sum_{j=1}^N \delta_0^{j-1} x_j ,\quad \gamma_{R} = \sum_{j=1}^N \delta_0^{N-j} p_j , \quad \delta_0\equiv -\frac{\mu}{J}, 
\end{equation} 
which mimic closely the MFs of the FKC when $|\mu|<J$: they are Hermitian, exponentially boundary-localized, are (almost) conserved, $[H,\gamma_L]= iJ(-\delta_0 )^N x_N$, $[H,\gamma_R]=-iJ \delta_0^N p_1$, and obey $[\gamma_L,\gamma_R]=iN\delta_0^N$, so they can be normalized and combined to form a split bosonic mode. However, the system is either dynamically unstable, or on the cusp of instability.

One can circumvent the dynamical instability of the BKC by allowing for dissipation. For Model 1 $(J=\Delta)$, the modes 
\begin{equation*}
{\gamma}_L^\text{c} \equiv \sum_{j=1}^N \delta_-^{j-1} x_j,\; {\gamma}_R^\text{c} \equiv \sum_{j=1}^N \delta^{N-j}_+ p_j ,\quad\delta_\pm \equiv -\frac{\mu\pm \kappa}{J},
\end{equation*}
satisfy $\mathcal{L}^\star(\gamma_L^\text{c}) = -J\delta_-^N x_N$ and  $ \mathcal{L}^\star(\gamma_R^\text{c}) = J\delta_+^N p_1$. Hence, $\gamma_R^\text{c}$ ($\gamma_L^\text{c}$) provides a right-(left-)localized approximate ZM whenever $|\delta_-|<1$ ($|\delta_+|<1$), that is, \emph{precisely when} the $x$($p$)-quadrature rapidity band winds around the origin. These operators have the physical meaning of (approximately) conserved quantities, and coincide with Eq.\,\eqref{Deconbosoranas} as $\kappa\to0$. 

\begin{figure}[t]
\includegraphics[width=0.85\columnwidth]{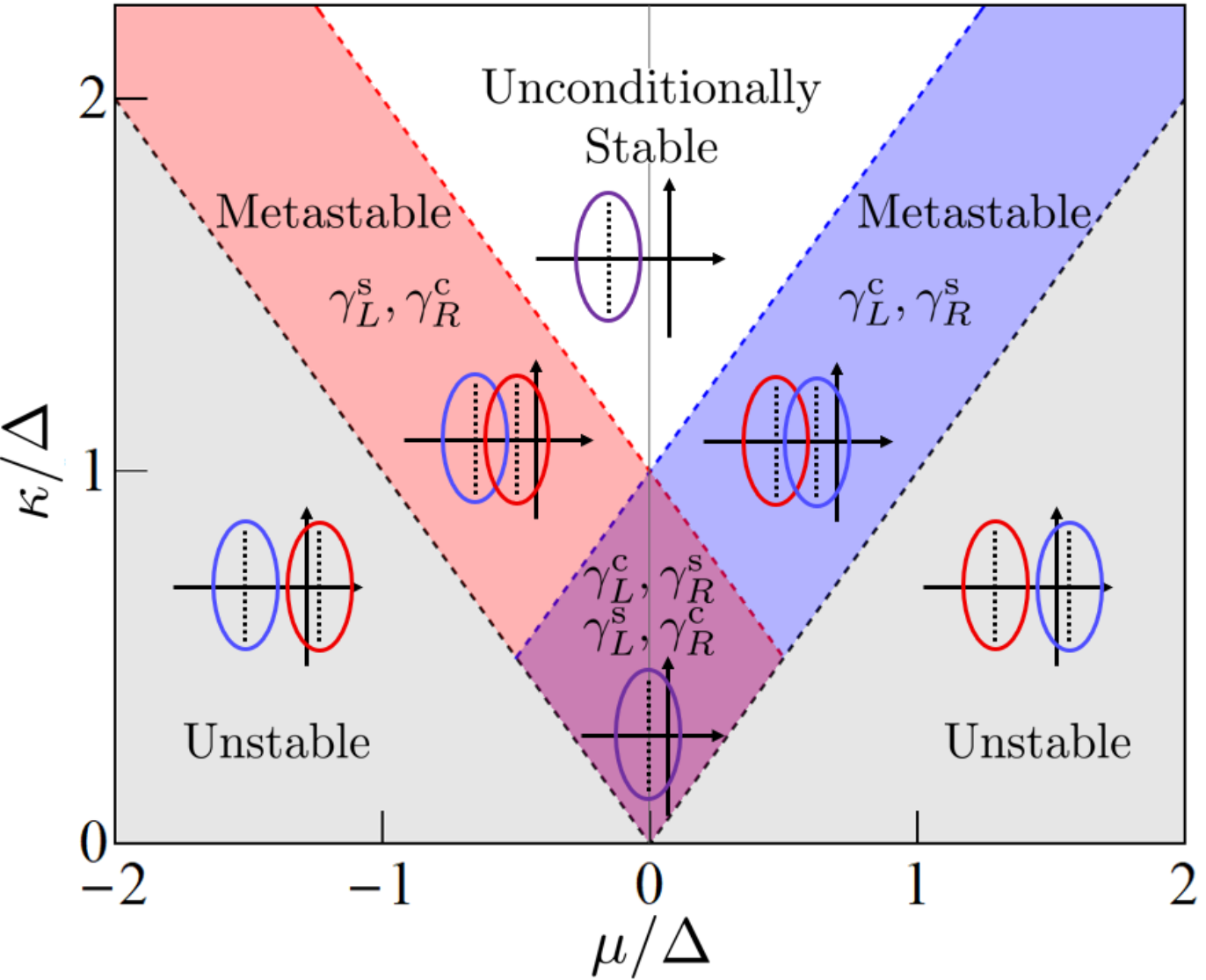}
\vspace*{-3mm}
\caption{(Color online) Topological phase diagram for Model 1. The blue (red) strip is the region where the rapidity band corresponding to the $x$ ($p$) quadrature winds around the origin (exemplified by the inset example spectra plotted in the complex plane; the black dashed line are the OBC spectrum). In each such strip, the system hosts a pair of MBs, whereas a total of four MBs exist in the central (purple) region. The diagram is insensitive to changes in $J\geq\Delta$.}
\label{f3_bkcpd}
\end{figure}

Noting that the system is stable whenever $|\mu|-\kappa<0$, it appears that we achieved our goal of finding MBs. However, there is a glaring discrepancy with the FKC: whenever $\kappa\neq 0$, it is \emph{not} generally possible to combine $\gamma^\text{c}_L$ and $\gamma^\text{c}_R$ into a split bosonic degree of freedom. This follows because (i) they can exist independently of each other (see also Fig.\,\ref{f3_bkcpd}); and (ii) they need not be canonically conjugate. For example, when $\mu=0$, $[\gamma_L^\text{c},\gamma_R^\text{c}] = 0$ for $N$ even. Instead, the conjugate modes required to form the split degree of freedom are
\begin{equation}
\label{sLsR}
\gamma_L^\text{s} \equiv \sum_{j=1}^N \delta_+^{j-1} x_j,\quad \gamma_R^\text{s} \equiv 
\sum_{j=1}^N \delta^{N-j}_- p_j ,
\end{equation} 
which satisfy $[\gamma_L^\text{c},\gamma_R^\text{s}] = iN\delta_-^{N-1}$ and $[\gamma_L^\text{s},\gamma_R^\text{c}]=iN\delta_+^{N-1}$. 

We propose that the pairs $(\gamma_L^\text{c},\gamma_R^\text{s})$ and $(\gamma_L^\text{s},\gamma_R^\text{c})$ are the \emph{bosonic analogues} of the MFs. There is a doubling of MB pairs with respect to the FKC. This is consistent with the breakdown of Noether's theorem for open systems \cite{Victor,Baum}, in the sense that conserved quantities ($c$) and symmetry generators ($s$) need no longer coincide (although a correspondence still exists for linear forms \cite{SM}). Explicitly, while in the Hamiltonian case the edge modes can each be thought of as supplying \emph{both} a conserved quantity and a generator of a U(1) symmetry, they now generically split into pairs of distinct ZMs and generators. The latter, given in Eq.\,\eqref{sLsR} obey
\begin{equation*}
\mathcal{L}^\star \left([\gamma^\text{s}_{L},A]\right)  - [\gamma^\text{s}_{L},\mathcal{L}^\star(A)] = - J\delta_{+}^N[x_N,A] , 
\quad \forall A, 
\end{equation*}
and similarly for $\gamma_R^\text{s}$, with $(\delta_+^N,x_N)\mapsto (\delta_-^N,-p_1)$. That is, if $U(\theta)\equiv e^{i\theta \gamma^\text{s}_{L,R}}, \theta \in {\mathbb R},$ the dynamics are invariant under the unitary action 
(or ``weak'' symmetry \cite{ProsenSpin,SM}) ${\cal L}(U(\theta)\rho U(\theta)^\dagger)= U(\theta){\cal L}(\rho) U(\theta)^\dagger,$ up to exponentially small corrections in system size.
 
To see the connection between MBs, topology, and pseudospectra  more generally, 
let us associate to each linear form $v$ a vector $\vec{v}\in\mathbb{C}^{2N}$ defined by $v=\vec{v}^{\,\dag} {\bm \tau}_3 {\Phi}$. A direct computation then yields \(\mathcal{L}^\star  (v) = w\), with $\vec{w} \equiv i {\bf \widetilde{G}}{\vec{v}}$ and \({\bf \widetilde{G}} \equiv {\bm \tau}_3 {\bf G}^\dag {\bm \tau}_3 \). From the definition of $\epsilon$-pseudospectra, one can verify that $\sigma_\epsilon(\mathbf{G}) = \sigma_\epsilon(\widetilde{\mathbf{G}})$. So, when the rapidity bands wind around $\lambda=0$, it follows that $0\in\sigma_\epsilon(\widetilde{\mathbf{G}})$, with $\epsilon$ exponentially small in $N$. The conserved MB is then associated with an $\epsilon$-pseudoeigenvector of $\widetilde{\mathbf{G}}$; the symmetry generator MB is formed from the corresponding $\epsilon$-pseudoeigenvector of $\mathbf{G}$ \cite{SM}. It may be possible for these pseudoeigenvectors to coincide, yielding `non-split' MBs. Importantly, MBs survive in the presence of disorder, thanks to the robustness of pseudospectra \cite{SM}. We call this phenomenon \emph{topological metastability} to contrast it with systems like Model 2, which can have vanishing winding number at zero (hence no MBs) when metastable. This topological metastability is forbidden in quadratic fermionic Lindbladians, due to the rapidity spectra being bounded to the left-half plane \cite{prosenfermspecthm}.

The approximate symmetries arising from topological metastability imply a degenerate quasi-steady state manifold. Let $\gamma^\text{s}$ be any linear approximate symmetry and consider the Weyl displacements \(\rho_\theta \equiv e^{i\theta\gamma^\text{s}}\rho_\text{ss}e^{-i\theta\gamma^\text{s}} \), with $\rho_\text{ss}$ the unique, Gaussian steady state under OBCs. In general, these quasi-steady states are arbitrarily long-lived. To see this, note that $\rho_\theta$ is also Gaussian, hence its lifetime is fully determined by those of the first and second moments, $\vec{m}_\theta(t)\equiv \tr[\Phi\rho_\theta(t)]$ and $\mathbf{Q}_\theta(t)\equiv \tr[Q\rho_\theta(t)]$. One can show that 
\begin{equation}
\label{mUB}
\frac{\norm{\vec{m}_\theta(t) -\vec{m}_\theta(0)}}{\norm{\vec{m}_\theta(0)}}
\leq \epsilon  t \!\sup_{\tau\in[0,t]}\norm{e^{-i\mathbf{G}\tau}}_2 
\leq \epsilon t \,e^{\Omega t}, 
\end{equation}
with $\Omega$ the size-independent constant of Eq. \eqref{tlin} and a similar bound holding for $\mathbf{Q}_\theta(t)$. Thus, the expectation of an arbitrary linear form $\langle \alpha \rangle_t/ [\norm{\vec{\alpha}}\norm{\vec{m}_\theta(0)}]$ remains within an accuracy $\delta$ from $\langle \alpha \rangle_0/[\norm{\vec{\alpha}}\norm{\vec{m}_\theta(0)}]$ for all times $t$ with $t e^{\Omega t} <\delta/\epsilon$ \cite{SM}. Since, for these systems, $\epsilon$ decays exponentially with $N$, this condition can be met for arbitrarily long $t$ (see also Fig.\,\ref{f2_expamplif}(c)-(d)). The quasi-steady states $\rho_\theta$ can be used to construct long-lived initial states with properties uncharacteristic of stable QBLs: e.g., convex combinations of the $\rho_\theta$'s are generically \emph{non-Gaussian}. 
Moreover, the existence of long-lived states with non-zero first moments is \textit{unique} to systems exhibiting topological metastability \cite{SM}.  

\begin{figure}[t!]
\includegraphics[width=\columnwidth]{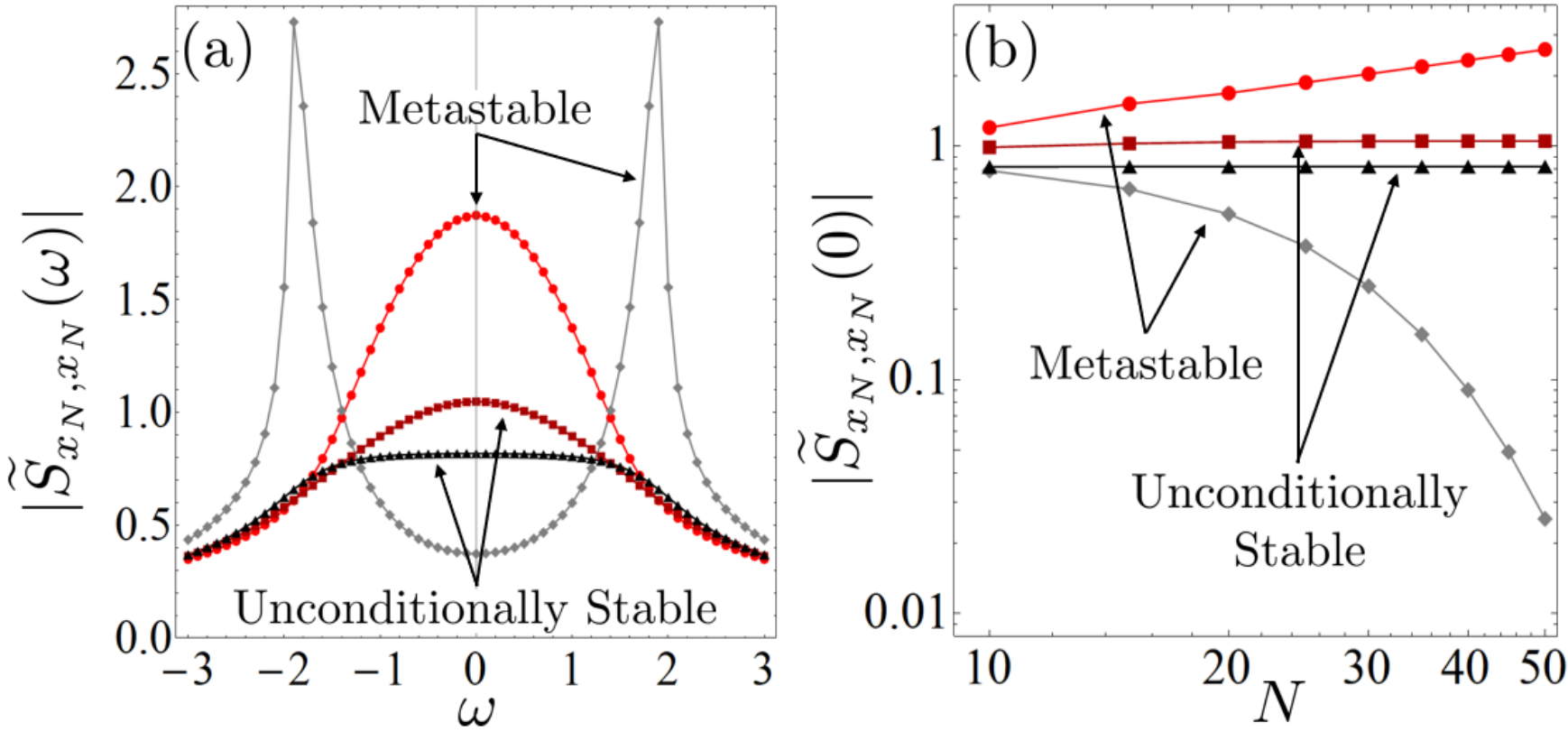}
\caption{(Color online) (a) Modulus of the normalized power spectra for the correlator of $\alpha=\beta=x_N$ in Model 1 (light red disks $\kappa/\Delta=0.6$ and dark red squares $\kappa/\Delta=1.4$) and Model 2 (gray diamonds $\kappa/\Delta=0.6$ and black triangles $\kappa/\Delta=1.4$), with $\Gamma=0.12$. In all cases $J=2$, $\Delta=0.5$, $N=25$. (b) Modulus of the zero-frequency component for  the same parameters in (a) as a function of $N$. }
\label{f4_PS}
\end{figure}

\emph{Towards experimental signatures of MBs.---} Consider the two-time correlators of two linear forms, $C_{\alpha,\beta}(t,\tau) \equiv \braket{\alpha(t+\tau)\beta^\dag(t)} = \tr [\alpha(\tau)\beta^\dag(0)\rho(t)],$ where the second equality relies on the quantum regression theorem \cite{Gardiner}. Because of Eq.\,\eqref{PhiEOM}, $C_{\alpha,\beta}(t,\tau) = \vec{\alpha}\,{}^\dag [\bm{\tau}_3 e^{-i\mathbf{G}\tau} \mathbf{Q}(t)\bm{\tau}_3]\vec{\beta}$. Taking the (one-sided) Fourier transform of this correlator as ${t\to\infty}$ yields the steady-state power spectrum, 
$$S_{\alpha,\beta}(\omega) \!= \vec{\alpha}\,{}^\dag[ \bm{\tau}_3 \bm{\chi}_N(\omega) \mathbf{Q}_\text{ss} \bm{\tau}_3]\vec{\beta},\,\,\,  
\bm{\chi}_N(\omega)\equiv i(\omega\bm{1}_{2N}-\mathbf{G})^{-1}.$$ Within input-output theory, $\bm{\chi}_N(\omega)$ is related to the susceptibility matrix that determines the transformation of an input field to an outgoing one, and a correspondence exists  between  topological properties of $\mathbf{G}$ and exponentially enhanced end-to-end amplification in driven-dissipative photonic lattices \cite{PorrasPRL,NunnenkampNat,NunnenkampDisorder,PorrasIO}. Utilizing pseudospectra, a non-zero winding of the bulk rapidity spectra around $i\omega$ mandates that $\norm{\bm{\chi}_N(\omega)}$ diverges with $N$, signaling amplification \cite{Clarification}.

While this amplification by itself is a manifestation of metastability, a \emph{necessary} condition for the existence of MBs is that $\norm{\bm{\chi}_N(0)}=\norm{\mathbf{G}^{-1}}$ diverges with $N$. To isolate the influence of MBs, we must account for possible divergent behavior of steady-state correlations, by computing the normalized power spectra \cite{NoriLEP} \(\widetilde{S}_{\alpha,\beta}(\omega) \equiv {S_{\alpha,\beta}(\omega)}/{C_{\alpha,\beta}(\infty,0)} \). Hence, we expect MBs to be diagnosed by a \emph{divergent zero-frequency peak} of certain steady-state normalized power spectra. We assess this conjecture in Fig.\,\ref{f4_PS}: a clear zero-frequency peak is seen for the BKC, in contrast to the MB-free model. Moreover, this peak appears to diverge as a power law in $N$. 

{\em Outlook.---} We have shown that Majorana bosons, despite being strictly forbidden for gapped free-boson Hamiltonians, can emerge in metastable QBLs. The appropriate phase diagram characterizes dynamical rather than thermodynamical stability. While there are natural ways to extend these ideas to higher dimension, a key question for characterizing SPT phases of free bosons is whether dynamical stability phase diagrams are dictated by some topological classification with an associated, dissipative bulk-boundary correspondence. The nature of our Majorana bosons suggests that the breakdown of Noether's theorem may play a role in answering this question. Experimental realizations of metastable QBLs offer another exciting venue for future research. Since the basic nonlinearities and dissipative couplings are readily available \cite{MattiasPRX,SCMIT,SCYale}, superconducting arrays appear especially well-positioned to possibly uncover Majorana bosons and probe their physics. 

\medskip
It is a pleasure to thank Joshuah Heath and Roberto Onofrio for a critical reading of the manuscript.  Work at Dartmouth was partially supported by the US National Science Foundation through Grants No. PHY-1620541 and PHY-2013974, the US Department of Energy, Office of Science, Office of Advanced Scientific Computing Research, under the Accelerated Research in Quantum Computing (ARQC) program, and the Constance and Walter Burke Special Projects Fund in Quantum Information Science. 

%\bibliography{BosoranaBIB}

\begin{thebibliography}{52}
\bibitem [{Cla()}]{Clarification}%
  \BibitemOpen
  \href@noop {} {}\bibinfo {note} {Explicitly, amplification has been linked to
  a non-trivial winding of $\bm{\chi}^{-1}_N(\omega)$ about the origin. From
  its definition, $\bm{\chi}^{-1}_N(\omega)$ winds around the origin whenever
  $-i\mathbf{G}$ winds around $i\omega$.}\BibitemShut {Stop}%
\end{thebibliography}


\end{document}